# Surveyor Gender Modifies Average Survey Responses – Evidence from Household Surveys in Four Sub-Saharan African Countries

Noah Haber[1,2], Paul Jake Robyn[3], Saidou Hamadou[3], Gervais Yama[4], Herve Hien[5], Davy Louvouezo[6] & Günther Fink[2,7]

1 Carolina Population Center, University of North Carolina at Chapel Hill, Chapel Hill, USA

2 Department of Global Health and Population, Harvard TH Chan School of Public Health, Boston, USA

3 Yaoundé Office, World Bank Group, Yaoundé, Cameroon

4 World Bank Group, Central African Republic

5 Centre MURAZ, Bobo-Dioulasso, Burkina Faso

6 Médecins d'Afrique, Brazzaville, Republic of the Congo

7 Swiss Tropical and Public Health Institute, University of Basel, Basel, Switzerland

Contact information

Noah Haber (nhaber@unc.edu)*

Paul Jake Robyn (probyn@worldbank.org)

Saidou Hamadou (saidouhtheo@yahoo.fr)

Gervais Yama (yama.gervais@gmail.com)

Herve Hien (hien_herve@hotmail.com)

Davy Louvouezo (dlouvouezo@gmail.com)

Günther Fink (guenther.fink@swisstph.ch)

* Corresponding author

# Abstract

## Background
While a large literature has highlighted the importance of wording and general interview settings for the collection of ideally unbiased survey data, relatively little is known regarding the influence of surveyor traits on respondent behavior. In this paper, we assess the extent to which survey gender modifies average survey responses in the context of health-focused household surveys.

## Methods
We pool data from four recent health-focused household surveys using both male and female surveyors: Burkina Faso (2014), Cameroon (2012), Central African Republic (2012), and Republic of Congo (2014). In all surveys, surveyors were pre-assigned to households based on an initial household listing. We compare responses given to male and female surveyors across three domains: household characteristics, child mortality and reproductive health. Multivariable regression models were used to estimate response differentials. Enumeration area fixed effects were used to remove spatial biases.

## Results
A total of 22,835 household surveys were analyzed. The proportion of interviews conducted by female interviews varied between 9 percent in Central African Republic and 52 percent in Cameroon. Female surveyor gender increased the odds of reporting asset ownership by 9.4% (OR 1.094, 95% CI: 1.024,1.169) and increased the odds of reporting a pregnancy-related event by 25% (OR 1.246 (95% CI: 1.12,1.393). Being interviewed by a woman increased the odds of respondents reporting a stillbirth by 29% (95% CI: 1.118,1.492), and the odds of reporting a miscarriage by 17% (95% CI: 1.072,1.284). Substantial heterogeneity in gender-specific reporting was found across the four countries. We did not find evidence that the gender of the participant modified the effect of surveyor gender for household items.

## Conclusions
All results presented in this paper suggest that surveyor gender is highly predictive of survey responses. For health surveys, female surveyors are likely to receive more accurate and consistent responses. More generally, social distance between interviewers and interviewees should be minimized in large scale surveys.

# Manuscript

## Introduction

Statistical inference from surveys typically relies on the assumption that observations received are an accurate, unbiased, and independently drawn reflection of factual "truth," or at least that misreporting is effectively random with respect to the research question. However, there are often differences between what is reported by respondents and their actual behaviors and statuses. Large bodies of literature and effort have been dedicated to theoretical development underlying misreporting, attempts to measure its effect, survey methodologies designed to minimize misreporting, and statistical methods to account for it after data have been collected. One key piece of conventional wisdom in survey best practices is that the gender of the surveyor should match the gender of the participant for live interviews to minimize biases due to social pressures.

Social distance theory predicts that reporting bias will increase with social distance between the interviewer and interviewee as well as with the relative importance of the question asked.[1] In practice, social distance may not always be clearly defined. However, the source of bias is fairly intuitive: the more the interviewers' perceived norms differ from participants' actual behavior or knowledge, the more likely subjects are to modify their responses to comply with perceived norms. While social distance was originally mostly defined across race, literature increasingly focused on gender differences.[2,3]

While differential responses to male and female interviewers seem likely conceptually, evidence on interviewer gender effects is limited and somewhat contradictory.[3] Interviewer gender effects appear to be best documented in cases of sexual behaviors[4-10] and gender-related political and social issues.[11-15] Effects were also observed in areas that were seemingly unrelated to sex and gender, including disclosure of loan amounts,[16] youth criminal activity,[17] and reporting concussion symptoms.[18] Findings are often inconsistent across geographic regions.[4] Interviewer gender effects even been observed in cases where the data collection took place without a live interviewer.[5,6,11,12,17]

In this paper, we use data from large population-representative household surveys in four African countries using both male and female interviewers to assess the degree to which survey gender predicts survey responses.

## Methods

### Data

We analyze four recent household surveys implemented by the World Bank in collaboration with ministries of health and local survey partners in Burkina Faso (2014), Cameroon (2012), Central African Republic (2012), and Republic of Congo (2014) as baselines surveys for current Results-based financing intervention projects conducted in these countries. All surveys used a two-stage cluster-randomized sampling procedure, first randomly selecting enumeration areas (EAs), and then randomly selecting households with women of ages 15-49 for the surveys. Respondents in this sample consisted only of adults (18+). For all surveys, a standardized household questionnaire was used. In all four surveys, selection of interviewers was left to local survey implementers, who recruited and trained surveyors specifically for these surveys.

Table 1: Survey Implementation Overview

| Country | Implementation Dates | Implementing Agencies | Number of households | # of surveyors | # of female surveyors |
|---|---|---|---|---|---|
| Burkina Faso | January-June, 2014 | Centre Muraz | 6,224 | 60 | 20 |
| Cameroon | March-June, 2012 | Institute of Training and Population Research (IFORD) | 3,874 | 32 | 16 |
| Central African Republic | September-December, 2012 | World Bank | 5,387 | 54 | 6 |
| Republic of the Congo | June-October, 2014 | Medicins d'Afrique | 7,442 | 55 | 12 |

Statistical Analysis

We start our analysis with descriptive statistics for all key variables by survey. To assess surveyor gender effects, we focused on 12 questions that were used in all four surveys. The first four questions pertained to household size and goods owned. These questions were asked to the head of household, with one response per household. Given that many households had a female head of households, we can separately analyze all four surveyor/interviewer gender combinations. The remaining eight questions pertain to reproductive health and birth histories. For these eight questions, only female respondents are included.

We use standard OLS regression models in our analysis. Given that surveyor team gender composition (27% female on average, ranging from 11% to 48%) varies across regions and villages, we include enumeration area fixed effects in our empirical models, and thus explore within-cluster variation in responses only. This model allows correct causal inference under the assumption that households within a given community are randomly assigned to interviewers, which seems plausible given that survey supervisors – who assign households to interviewers – generally have no information on households and simply assign households sequentially to staff based on the list of eligible households.

For binary variables, we used logistic regression models; for continuous variables, standard OLS models were used. All estimates include enumeration area clustered standard errors and confidence intervals. Country-specific fixed effects are directly absorbed by the lower level cluster fixed effects. To explore possible interaction effects between the genders of the surveyor and the participant we include dummy terms for all combinations in the analysis of the household variables (where respondent gender varies).

To explore heterogeneity across countries for all questions, we run separate country-specific regressions to estimate the impact of surveyor gender for each country.

We additionally use meta-analysis to estimate pooled effects across categories of dependent variables, referred to in the text as "meta-pooled" coefficients. We assume a random-effects model, where we are estimating an average effect across several independent effects, rather than assuming the existence of a "true" effect. In order to make coefficients more comparable across variables, we additionally used z-score, or a standard normalized variant of the dependent variables. Dependent variables were transformed into standardized outcomes by subtracting the mean and dividing by the standard deviation, yielding a z-score. This allows use of both continuous and binary variables together in meta-regression, whereas the logistic meta regression only allows binary outcomes variables.

All analysis and charts were generated using Stata 14.[19] Meta-analysis was performed using the *metan* package.[20] The full model equations and standardization formulas we estimate are shown in Appendix 1.

# Results

*Table 1: Descriptive statistics*

|  | All countries | Burkina Faso | Cameroon | Central African Republic | Republic of the Congo |
|---|---|---|---|---|---|
| Female | 57% (0.5) | 56% (0.5) | 60% (0.49) | 56% (0.5) | 56% (0.5) |
| Age at interview | 33.3 (12.6) | 32.9 (11.1) | 34.7 (14.5) | 31.6 (11.3) | 34 (13.4) |
| Household size | 5.37 (2.33) | 5.4 (2.42) | 5.66 (2.38) | 5.68 (2.46) | 4.96 (2.03) |
| Household owns cell phone | 64% (0.48) | 86% (0.35) | 68% (0.46) | 15% (0.36) | 80% (0.4) |
| Household owns television | 21% (0.41) | 8% (0.27) | 37% (0.48) | 2% (0.15) | 39% (0.49) |
| Household owns car | 1% (0.12) | 0% (0.05) | 5% (0.21) | 0% (0.06) | 2% (0.13) |
| Woman has had a live birth | 95% (0.22) | 97% (0.16) | 96% (0.2) | 92% (0.26) | 94% (0.24) |
| At least one live birth died | 23% (0.42) | 21% (0.41) | 24% (0.42) | 34% (0.47) | 15% (0.36) |
| Has had a stillbirth | 5% (0.22) | 2% (0.14) | 6% (0.23) | 7% (0.26) | 5% (0.23) |
| Ever had a miscarriage | 13% (0.34) | 8% (0.27) | 16% (0.36) | 9% (0.28) | 20% (0.4) |
| Currently pregnant | 18% (0.38) | 13% (0.34) | 20% (0.4) | 23% (0.42) | 18% (0.38) |
| Current pregnancy was undesired | 35% (0.48) | 18% (0.39) | 35% (0.48) | 34% (0.48) | 46% (0.5) |
| Currently has a regular sex partner | 83% (0.38) | 92% (0.27) | 64% (0.48) | 87% (0.34) | 81% (0.39) |
| Currently uses contraceptives | 26% (0.44) | 12% (0.33) | 52% (0.5) | 18% (0.39) | 39% (0.49) |
| Number of individuals | 51,423 | 14,209 | 9,369 | 12,124 | 15,721 |
| Number of households | 22,835 | 6,213 | 3,869 | 5,358 | 7,395 |
| Number of clusters | 977 | 417 | 242 | 9 | 309 |
| Number of interviewers | 199 | 57 | 33 | 54 | 55 |
| Number of female interviewers | 54 | 20 | 16 | 6 | 12 |
| % of interviews by female interviewers | 31% | 39% | 52% | 9% | 27% |

*Descriptive statistics includes all interviews in the dataset, with both male and female surveyors. Standard deviations in parentheses*

The four surveys captured a total of 51,423 individuals from 22,835 households, as shown in Table 1. 57% of respondents were women, with a mean age of 33. There were 54 female interviewers out of the 199 total interviewers across the four surveys, with 27% of the interviewers being female, and 31% of interviews performed by female interviewers.

*Table 2: Associations between surveyor gender and reported household characteristics*

*Panel a: Basic regressions*

| Model | Household size OLS | Household owns cell phone Logistic (odds ratio) | Household owns television Logistic (odds ratio) | Household owns car Logistic (odds ratio) |
|---|---|---|---|---|
| Female interviewer | 0.0453 | 1.098** | 1.072 | 1.210 |
|  | [-0.0441,0.135] | [1.005,1.201] | [0.965,1.191] | [0.924,1.584] |
| EA fixed effects | YES | YES | YES | YES |
| Observations | 21,967 | 19,240 | 17,548 | 8,524 |
| R-squared | 0.11 | 0.000283 | 0.000173 | 0.00111 |

*Panel b: Participant gender interaction*

|  | Household size | Household owns cell phone | Household owns television | Household owns car |
|---|---|---|---|---|

| Model | OLS | Logistic (odds ratio) | Logistic (odds ratio) | Logistic (odds ratio) |
|---|---|---|---|---|
| Female interviewer | -0.00208 | 1.110 | 1.103 | 1.564* |
|  | [-0.134,0.130] | [0.966,1.275] | [0.923,1.319] | [0.919,2.663] |
| Female respondent | -0.475*** | 0.856*** | 0.893* | 1.356 |
|  | [-0.551,-0.399] | [0.777,0.943] | [0.784,1.017] | [0.920,2.000] |
| Female interviewer * Female respondent | 0.0937 | 0.988 | 0.962 | 0.718 |
|  | [-0.0455,0.233] | [0.831,1.175] | [0.777,1.191] | [0.400,1.288] |
| EA fixed effects | YES | YES | YES | YES |
| Observations | 21,967 | 19,240 | 17,548 | 8,524 |
| R-squared | 0.12 | 0.00121 | 0.00072 | 0.00257 |

*Notes: All specifications control for enumeration area fixed effects. 95% confidence intervals, shown in brackets, are clustered at the enumeration area level. Observations are one observation per household.*

As shown in Table 2, average differences in reported household size and owned goods are small and generally not statistically significant. However, all effects measured for household size and goods are positive. When all four variables in this category are converted to z-scores and combined using meta-analysis, we find that female interviewers have a significant increase average in reporting of household size and goods, with 1.094 (95% CI: 1.024,1.169) times the odds of reporting household goods for the three binary variables in this category. On average, households with a female respondent (mostly with a female head of household) were smaller and marginally poorer. We did not find evidence for differential gender effects in households with female respondents in any individual question, nor were significant effects observed when pooling questions together in meta-regression.

*Table 3: Surveyor gender and reported pregnancy history events*

| Model | Woman has had a live birth<br>Logistic (odds ratio) | At least one live birth died<br>Logistic (odds ratio) | Has had a stillbirth<br>Logistic (odds ratio) | Ever had a miscarriage<br>Logistic (odds ratio) |
|---|---|---|---|---|
| Female interviewer | 1.505*** | 1.126*** | 1.291*** | 1.173*** |
|  | [1.292,1.753] | [1.044,1.214] | [1.118,1.492] | [1.072,1.284] |
| EA fixed effects | YES | YES | YES | YES |
| Observations | 15,351 | 23,712 | 15,688 | 21,992 |
| R-squared | 0.00393 | 0.00044 | 0.00167 | 0.000808 |

*Notes: All specifications control for enumeration area fixed effects. 95% confidence intervals, shown in brackets, are clustered at the enumeration area level. Observations are one observation per household. All reproductive health questions were answered by female respondents only.*

The odds of reporting each of the four questions pertaining to pregnancy events increased significantly when asked by a female surveyor, as shown in Table 3. When interviewed by a female, the odds of reporting a live birth increased 51% (OR 1.505, 95% CI: 1.292,1.753), the odds of reporting at least one live birth increased 13% (OR 1.126, 95% CI: 1.044,1.214), the odds of reporting a stillbirth increased 29% (OR 1.291, 95% CI: 1.118,1.492), and the odds of reporting a miscarriage increased 17% (OR 1.173, 95% CI: 1.072,1.284). When pooled together in meta-regression, the mean impact of having a female surveyor an increase in the odds of reporting a pregnancy-related event was 25% greater (OR 1.246, 95% CI: 1.115,1.393).

*Table 4: Associations between surveyor gender and reproductive health outcomes*

| Model | Currently pregnant<br>Logistic (odds ratio) | Current pregnancy was undesired<br>Logistic (odds ratio) | Currently has a regular sex partner<br>Logistic (odds ratio) | Currently uses contraceptives<br>Logistic (odds ratio) |
|---|---|---|---|---|
| Female interviewer | 1.041 | 0.837** | 1.062 | 1.210*** |
| | [0.961,1.127] | [0.703,0.997] | [0.971,1.160] | [1.111,1.317] |
| EA fixed effects | YES | YES | YES | YES |
| Observations | 24,122 | 3,695 | 19,913 | 18,321 |
| R-squared | 0.0000486 | 0.00112 | 0.000112 | 0.00122 |

*Notes: All specifications control for enumeration area fixed effects. 95% confidence intervals, shown in brackets, are clustered at the enumeration area level. Observations are one observation per household. All reproductive health questions were answered by female respondents only.*

Table 4 shows the main regression results on reproductive health. No significant difference was observed for being currently pregnant nor for currently having a regular sex partner. However, female surveyor yielded a 16% decrease in the odds of reporting that the current pregnancy was undesired (OR: 0.84, 95% CI: 0.703,0.997) and a 21% increase in the odds of reporting currently yielding. Meta-analysis did not yield significant differences when pooled (OR 1.048, 95% CI: 0.935,1.174).

*Figure 1: Coefficient sizes for main fixed effect regression expressed as odds ratios*

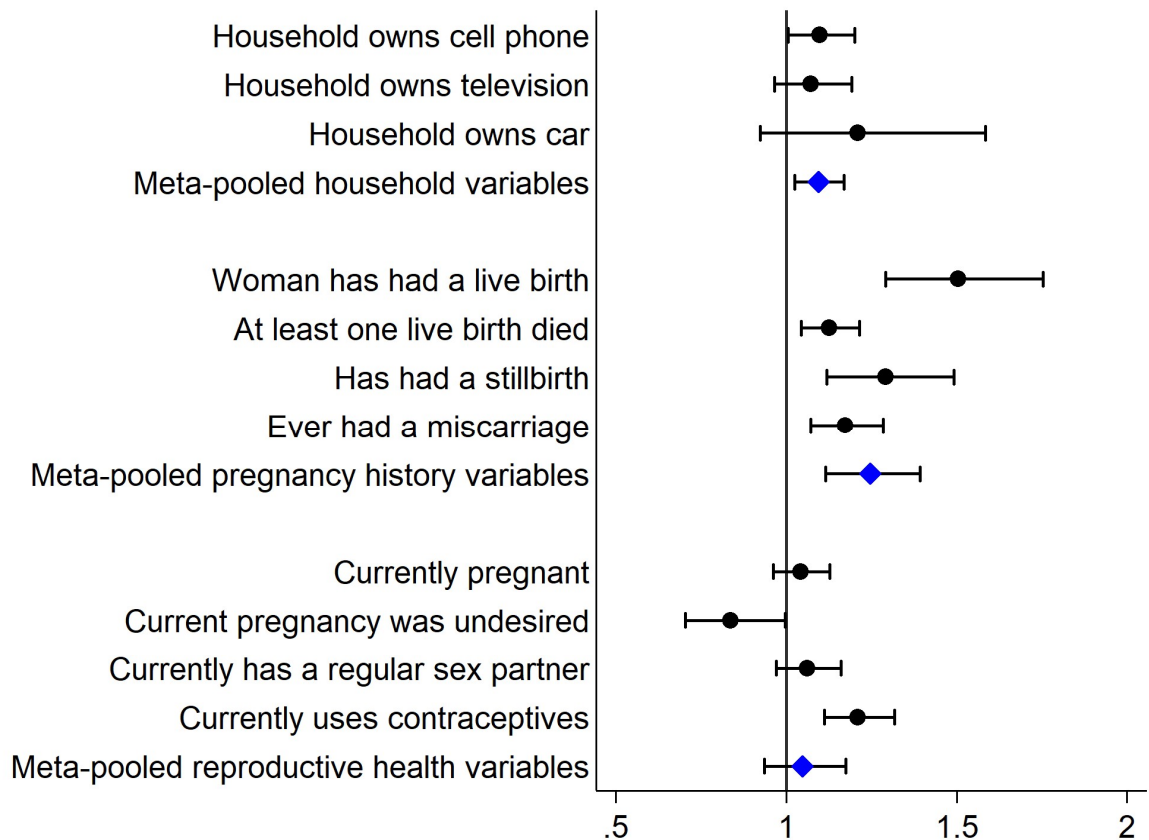

*Notes: The points represent the odds ratio coefficient sizes from the logistic fixed-effects regression shown in the methods section. The 95% confidence interval is shown by the bar width, using enumeration area-clustered standard errors.*

*Figure 2: Coefficient sizes for main fixed effect regression expressed as standardized dependent variables*

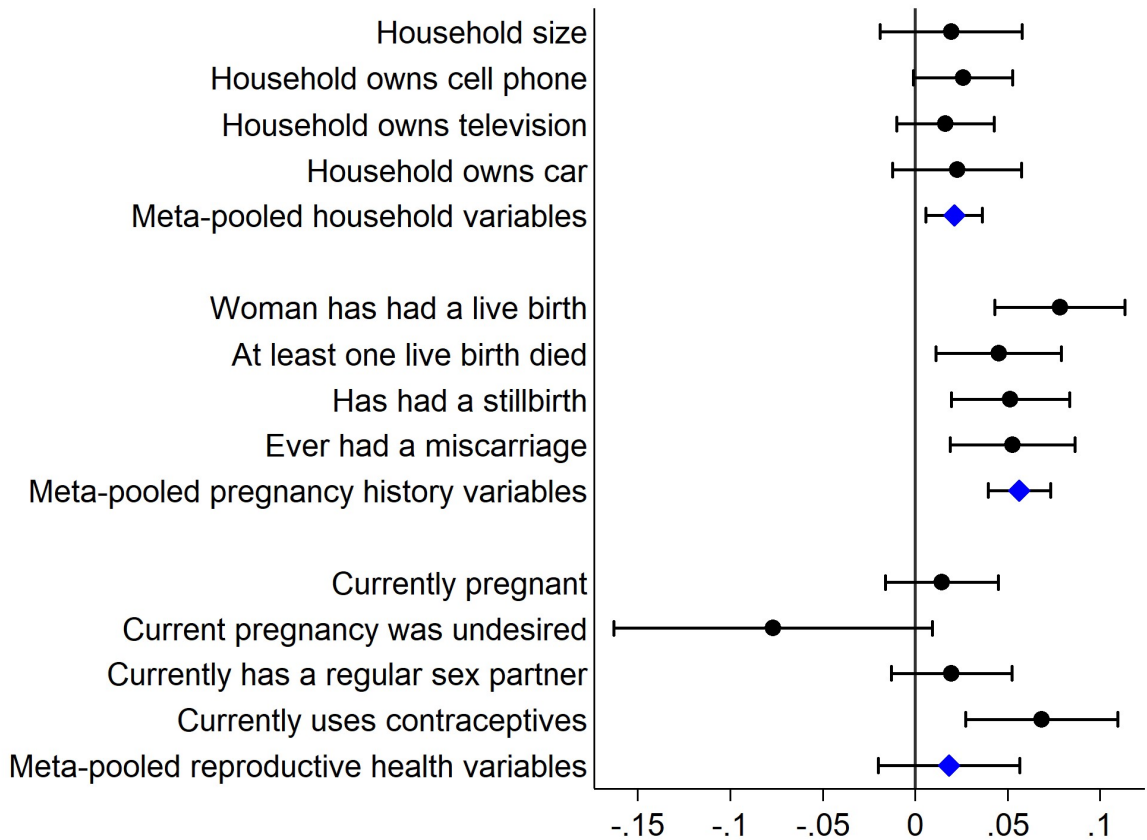

*Notes: The points represent the z-score standardized coefficient sizes from the fixed-effects regression shown in the methods section, using the standardized dependent variables. The 95% confidence interval is shown by the bar width, using enumeration area-clustered standard errors.*

The results for all binary dependent variables expressed as odds ratios in Figure 1, with from the standardized regressions are shown in Figure 2. As seen above, the questions pertaining to reproductive health are the most likely to be impacted by gender of the interviewer, while household factors and birth histories yielded more mixed results. Pooling the variables by category, we see that having a female surveyor increased the odds of reporting household goods and pregnancy events, but did not have a significant increase in the pooled odds of reproductive health outcomes. Whether the current pregnancy was desired or undesired is an outlier among these questions, as it is the only question in which the impact of having a female surveyor was reported fewer events.

The effect of interviewer sex across countries is highly heterogeneous across countries, as shown in Appendix 2. In two cases, household size and having a regular sex partner, different countries have opposite and independently significant effects. One four questions out of the 13 total questions had effects with the same sign across all four countries, pertaining to ownership of a cell phone, ownership of a car, ever having a miscarriage, and use of contraception.

Discussion

This study found that surveyor gender had significant, but inconsistent effects on survey-elicited data from four African countries. We found that household heads were most likely to report higher ownership of household goods, and that women were more likely to report pregnancy-related events when asked by a female interviewer, with inconsistent results for questions regarding sexual behavior. However, we did not find evidence that differences in reporting of household goods by surveyor gender interacted with the gender of the participant, suggesting while the gender of the surveyor may impact results, gender matching itself did not. Observed differences were likely to be heterogenous across survey countries.

The conventional wisdom that participants are more likely to respond truthfully, particularly for sensitive events, appears to hold in the study, albeit inconsistently. While *a priori* we might have expected that the impact of interviewer gender would have been strongest for questions regarding reproductive health and pregnancy histories, we did not observe consistent differences for reproductive health. Instead, we found interviewer gender most strongly impacted reporting of household economic goods and pregnancy histories.

The largest limitations of this study are ones of generalizability. These surveys were conducted in four African countries in concert with World Bank-supported development evaluations. We found evidence of between-country heterogeneity even in our highly select sample of countries. The selection of questions in this survey have further limited generalizability. These questions were generated to be comparable with other multi-national development surveys, in particular the Demographic and Health Survey. However, these questions are not a complete or strongly generalizable list of questions which could be asked with regard to household goods, pregnancy histories, and reproductive health. Finally, while we assume that the gender of assigned interviewers is as good as random in this case, it is plausible that interviewer gender may have impacted consent to participate in the study.

This study largely agrees with the idea that the gender of the interviewer can have important impact on survey elicited data, but does not find evidence for the benefit of gender matching. This suggests that researchers take care to understand measurement error that may be occurring due to interaction between interviewer and participant. In some cases, the impact of surveyor gender can be addressed through statistical means, particularly where its impact is well-understood. In others, it may simply be a limitation of the data and analysis.

*Appendix 1: Formal estimation procedures*

Standard OLS regression:
$$r_{ic} = \beta_0 + \beta_1 SurveyorFemale_i + \delta_c + \epsilon_{ic} \quad (1.1)$$

Where $r_{ic}$ is the response given by the interviewee, $SurveyorFemale_i$ is an indicator for whether or not the questions were asked by a female surveyor, and $\delta_c$ are cluster (enumeration-area) fixed effects.

Interaction with participant gender:
$$r_{ic} = \beta_0 + \beta_1 SurveyorFemale_i + \beta_2 ParticipantFemale_i + \beta_3 SurveyorFemale_i * ParticipantFemale_i + \delta_c + \epsilon_{ic} \quad (1.2)$$

Where $ParticipantFemale_i$ is a dummy indicator for whether the participant is female, and the interaction term $SurveyorFemale_i * ParticipantFemale_i$ is the additional effect of the interviewer and the participant both being female, in addition to the effects of each being true alone.

Standardization formula for dependent variables (z-score):
$$std(r_{ic}) = [r_{ic} - mean(R)]/stdev(R)$$

Where $r_{ic}$ is the value of the dependent variable as observed, and $R$ is the vector of all values of that variable.

Standardized OLS dependent variable OLS regression:
$$std(r_{ic}) = \beta_0 + \beta_1 SurveyorFemale_i + \delta_c + \epsilon_{ic} \quad (1.3)$$

*Appendix 2: Fixed effects regressions for each country*

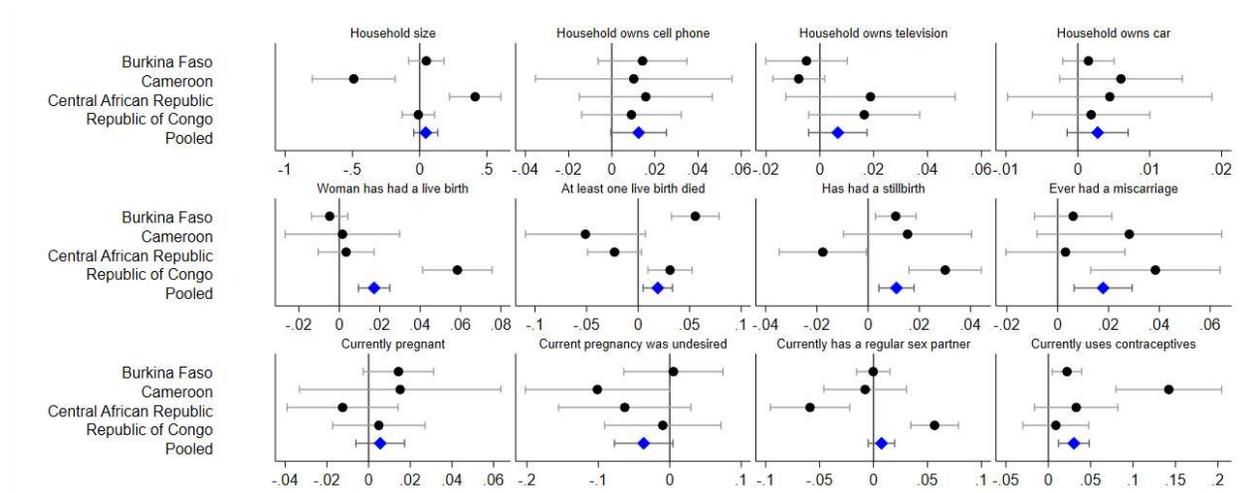

Notes: The points represent the coefficient sizes from the fixed-effects regression shown in the methods section. Coefficients are not normalized. The 95% confidence interval is shown by the bar width, using enumeration area -clustered standard errors for the pooled analyses

# Works cited

1. Williams Jr JA. Interviewer- respondent interaction: A study of bias in the information interview. *Sociometry : a journal of research in social psychology* 1964; **27**(3): 338-52.
2. Landis JR, Sullivan D, Sheley J. Feminist Attitudes as Related to Sex of the Interviewer. *The Pacific Sociological Review* 1973; **16**(3): 305-14.
3. Davis RE, Couper MP, Janz NK, Caldwell CH, Resnicow K. Interviewer effects in public health surveys. *Health Education Research* 2009; **25**(1): 14-26.
4. Becker S, Feyisetan K, Makinwa-Adebusoye P. The effect of the sex of interviewers on the quality of data in a Nigerian family planning questionnaire. *Studies in Family Planning* 1995; **26**(4): 233.
5. Catania JA, Binson D, Canchola J, Pollack LM, Hauck W, Coates TJ. Effects of interviewer gender, interviewer choice, and item wording on responses to questions concerning sexual behavior. *Public Opinion Quarterly* 1996; **60**(3): 345.
6. Fuchs M. Gender-of- Interviewer Effects in a Video-Enhanced Web Survey: Results from a Randomized Field Experiment. *Social Psychology* 2009; **40**(1): 37-42.
7. Agula J, Barrett JB, Tobi H. The Other Side of Rapport: Data Collection Mode and Interviewer Gender Effects on Sexual Health Reporting in Ghana. *African journal of reproductive health* 2015; **19**(3): 111-7.
8. Chun H, Tavarez MI, Dann GE, Anastario MP. Interviewer gender and self-reported sexual behavior and mental health among male military personnel. *International journal of public health* 2011; **56**(2): 225-9.
9. Abramson PR, Handschumacher IW. Experimenter effects on responses to double-entendre words. *Journal of personality assessment* 1978; **42**(6): 592-6.
10. Lamb ME, Garretson ME. The effects of interviewer gender and child gender on the informativeness of alleged child sexual abuse victims in forensic interviews. *Law and human behavior* 2003; **27**(2): 157-71.
11. Huddy L, Billig J, Bracciodieta J, Hoeffler L, Moynihan P, Pugliani P. The Effect of Interviewer Gender on the Survey Response. *Political Behavior* 1997; **19**(3): 197-220.
12. Kane E, Macaulay L. Interviewer Gender and Gender Attitudes. *Public Opinion Quarterly* 1993; **57**(1): 1-28.
13. Liu M, Stainback K. Interviewer Gender Effects on Survey Responses to Marriage-Related Questions. *Public Opinion Quarterly* 2013; **77**(2): 606-18.
14. Padfield A, Procter I. The effect of interviewer's gender on the interviewing process: A comparative enquiry. *Sociology-The Journal Of The British Sociological Association* 1996; **30**(2): 355-66.
15. Fry RPW, et al. Interviewing for Sexual Abuse: Reliability and Effect of Interviewer Gender. *Child Abuse & Neglect: The International Journal* 1996; **20**(8): 725-29.
16. Karlan D, Zinman J. Lying about borrowing. *Journal Of The European Economic Association* 2008; **6**(2-3): 510-21.
17. Dykema J, Diloreto K, Price J, White E, Schaeffer NC. ACASI Gender-of- Interviewer Voice Effects on Reports to Questions about Sensitive Behaviors Among Young Adults. *Public Opin Q* 2012; **76**(2): 311-25.
18. Krol AL, Mrazik M, Naidu D, Brooks BL, Iverson GL. Assessment of symptoms in a concussion management programme: method influences outcome. *Brain injury* 2011; **25**(13-14): 1300-5.
19. StataCorp. Stata Statistical Software: Release 14. College Station, TX: StataCorp LP; 2015.
20. Ross H, Mike B, Jon D, et al. METAN: Stata module for fixed and random effects meta-analysis. S456798 ed: Boston College Department of Economics; 2006.